# MEASUREMENT OF THE POLYTROPIC INDEX DURING SOLAR CORONAL RAIN USING A DIAGRAM OF THE ELECTRON DENSITY DISTRIBUTION AS A FUNCTION OF ELECTRON TEMPERATURE

**Z. M. Vashalomidze,[1] T. V. Zaqarashvili,[1,2,3] and V. D. Kukhianidze[1]**

*A differential emission measure (DEM) method is used to evaluate the relationship of the electron density and temperature before and after a coronal rain event during an active sun over the period from 20:10 UT on October 6 to 02:10 on October 7, 2011. Observational data were obtained from SDO/AIA for six different extreme ultraviolet (EUV) spectral lines. 240 different coronal loops were analyzed during this time interval, and the average electron density and temperature were obtained using 171 Å (Fe IX) and 193 Å (Fe XII) filters. The relationship between the density and temperature made it possible to estimate the polytropic index in the solar corona before and after the coronal rain. The polytropic index after termination of the coronal rain was estimated to be $\gamma = 1.3\pm0.06$, Which shows usual thermodynamic properties of study-state coronal plasma. The polytropic index at the time of onset of the coronal rain was, however, estimated to be $\gamma = 2.1\pm0.11$, which indicates an unstable thermodynamic process, i.e., a thermal instability. It is suggested that the coronal rain is the result of an unstable process, and the coronal plasma returns its stable state after the rain.*

[1] E. K. Kharadze Abastumani Astrophysical Observatory, Ilia State University, Tbilisi, Georgia; e-mail: zurab.vashalomidze.1@iliauni.edu.ge, teimuraz.zaqarashvili@uni-graz.at, vaso@iliauni.edu.ge
[2] Space Research Institute, Austrian Academy of Sciences, Schmiedlstrasse 6, 8042 Graz, Austria
[3] IGAM-Kanzelhohe Observatory, Institute of Physics, University of Graz, Universitatsplatz 5, 8010 Graz, Austria

## 1. Introduction

Coronal rain consists of cold, dense plasma bolobs falling along solar coronal loops in the direction of their support points. The blobs form rapidly in the coronal loops owing to a thermal instability through catastrophic cooling during which radiative losses overcome heat input [1-7]. Murawski, et al. [8], suggest that coronal rain may be formed by entropy modes at zero points during nanoflares in the corona. The entropy mode is characterized by enhancement of the local density and a temperature drop. Reconnection outflows carry cold bolbs from the null points, thereby forming the coronal rain.

In analytic studies of the solar corona the polytropic index or the ratio of the specific heats is the key parameter in thermal physics or thermodynamics. When heat transfer does not take place between the loops and the surrounding medium, the polytropic index transforms to the adiabatic index. For quasiadiabatic processes the polytropic index is in the range $\gamma \in [1; 5/3]$. During thermal instability or related processes, however, the polytropic index can exceed $\gamma = 5/3$. A first estimate of the adiabatic index in the solar corona was made by van Doorsselaere, at al. [9], using data from EIS/Hinode. Using a time dependent spectroscopic method, they detected slow magnetoacoustic oscillations in the electron density. The standard relationship between the change in the temperature and density yielded an estimate for the adiabatic index.

Here we use a differential emission measure (DEM) method to study the formation and thermal evolution of a system of coronal loops.

A first DEM method for studying the solar coronal plasma was developed by Fludra and Sylwester [10]. A few years later, Brosius, et al. [11], corrected this method with an iterative fitting with cubic spline functions. From time-to-time, other methods have been used, such as: a Monte-Carlo method with Markov chains [12], a single-Gaussian approximation [13-22], and multiple-Gaussian functions [14]. An automated direct approximation for a single-Gaussian DEM distribution with an automatic code for detection of a loop with high statistics and thermal analysis of coronal loops has been developed in Ref. 22. We used the DEM code proposed there [22] and accessible at SSW (SolarSoftWare). This automated data analysis was developed for recovery of electron temperatures and densities and some geometric parameters of coronal loops from the observed movement of the plasma in six EUV channels of the SDO (Solar dynamic observatory).

## 2. Data observation and analysis

The coronal rain studied here was observed with the SDO/AIA on October 6, 2011, on the East limb margin of the sun near the active region AR 11312 for 6 hours between 20:10 UT on October 6 and 02:10 UT on October 7. We used six extreme ultraviolet narrow band filters at 171 Å, 193 Å, 211 Å, 94 Å, 335 Å, and 131 Å with an effective spatial resolution of 1".6 [23]. These lines correspond to a coronal temperature from $10^{5.8}$ K to $10^{7.2}$ K. Satellite data were downloaded by the standard SSW packet and analyzed using the automated instruments for analysis AIA, which includes a determination of DEM temperature and density distributions [22].

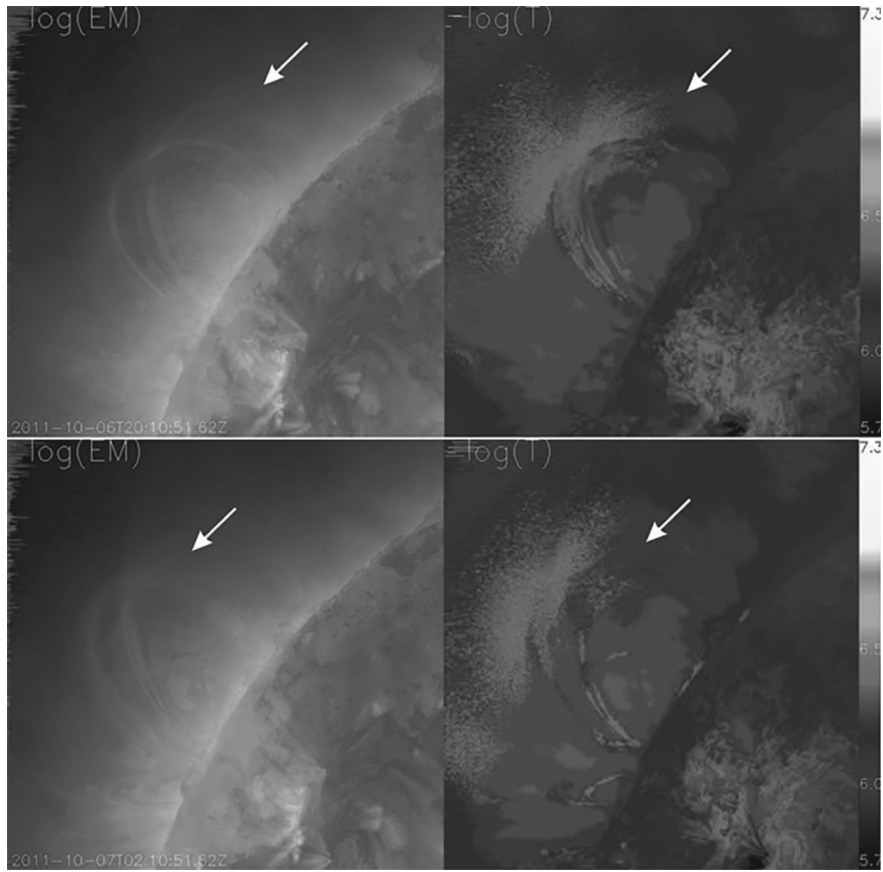

Fig. 1. Temperature and emission maps at the beginning and end of the coronal rain. Maps of the emission measure are shown in the top and bottom left frames and the temperature maps are shown on the top and bottom right frames. At the onset of the coronal rain, the emission chart in the top frame indicates (the white arrow in the top left panel) a high temperature at the vertex of the loop. At the end of the coronal rain the loops have almost disappeared in the emission maps (bottom left panel), but by the end of the event, the temperature is, as before, high (bottom right panel).

The formation of coronal rain is associated with rapid cooling of coronal loops [24]. Maps of the emission and temperature are produced by the DEM code [14]. Maps of the active region were taken on October 6, 2011, at 20:10 UT (upper left frame in Fig. 1). Coronal loops can be seen clearly in logarithmic charts of the emission at 20:10 UT (indicated by a white arrow), but at the end of the coronal rain (in the bottom left panel) by 02:11 UT, they gradually fade. The temperature maps derived from the six AIA wavelengths cover a range of $\log(T) \approx 6.2 - 6.4$ with the temperature scale indicated by a vertical strip; they indicate a higher temperature at the vertex of the coronal loop (Fig. 1, upper right frame). At the end of the coronal rain the temperature is again high, with $\log(T) \approx 6.2 - 6.4$ (Fig. 1, lower right frame).

The program for measuring the temperature and the emission makes an approximation of single-Gaussian DEM distributions, determines the peak emission, maximum temperature, and temperature for each pixel [22]. The overall

DEM distribution for the active region AR 11312 has a main peak at a temperature of $\log(T) \approx 6.2$ K.

### 3. Automatic determination of a coronal loop and temperature analysis

A coronal rain is related to the individual structure of a coronal loop, so that after processing of the active region AR 11312 we analyzed the individual coronal loops. For automatic detection and analysis of the coronal loops, we used an Oriented Coronal CUrved Loop Tracing (OCCULT) code [11-12]. The code includes automatic detection, extraction and tracking of curve linear features which are aligned on Solar EUV and SXR images [22]. At the beginning of the coronal rain, OCCULT automatically detected 240 loop-shaped objects. Most of the loops showed up in the 171 Å (Fe IX), 193 Å (Fe XII) and 211 Å (Fe XIV) lines. Over the next four hours, however, the number of loops decreased significantly; this is evidently related to a reduced temperature of the coronal loops. It was not possible to track the lost (cooled) loops using the DEM code if they showed up in the 304 Å line, which corresponds to a lower temperature.

We used the following control parameters: minimum radius of curvature of the loops $r_{min}$ = 30 pixels, typical half width of the loops $\omega$ = 4 pixels, threshold level in standard deviations of the flow $n_{sig}$ = 1.0, and maximum filling coefficient for the tracing structure $q_{fill}$ = 0.35. The tracing structures of the combined cycle for all six wavelengths

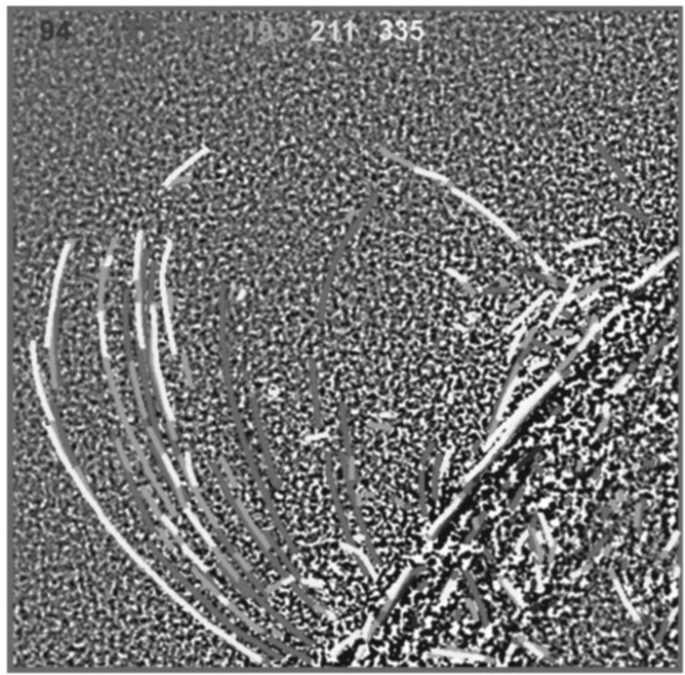

Fig. 2. Coronal loops determined automatically by the OCCULT code for six wavelengths. The different shadings represent different EUV channels (the wavelengths are indicated in the upper part of the image).

are shown in Fig. 2.

The OCCULT code with automated setting of the DEM provides automatically determined measures of the emission and temperature of the segments of a coronal loop [19,20,22]. The analysis was done for the 240 automatically detected coronal loops, which were broken down into 1355 subsegments. An example is shown for an AIA wavelength of 171 Å (Fe IX) in Fig. 3, which shows the automatically detected cycle with profiles of the transverse cross section of the flow along a coronal loop (upper left frame of Fig. 3) and with single-Gaussian DEM parameter approximations. This cycle includes the electron temperature along the coronal loop and the Gaussian temperature width indicated as error intervals for the temperature values, the electron density along the loop, the loop width, and an evaluation of the approximation (Fig.3, on the right from the middle to the bottom frame). As noted above, the loops show up most clearly in the 171 Å (Fe IX) filter [22].

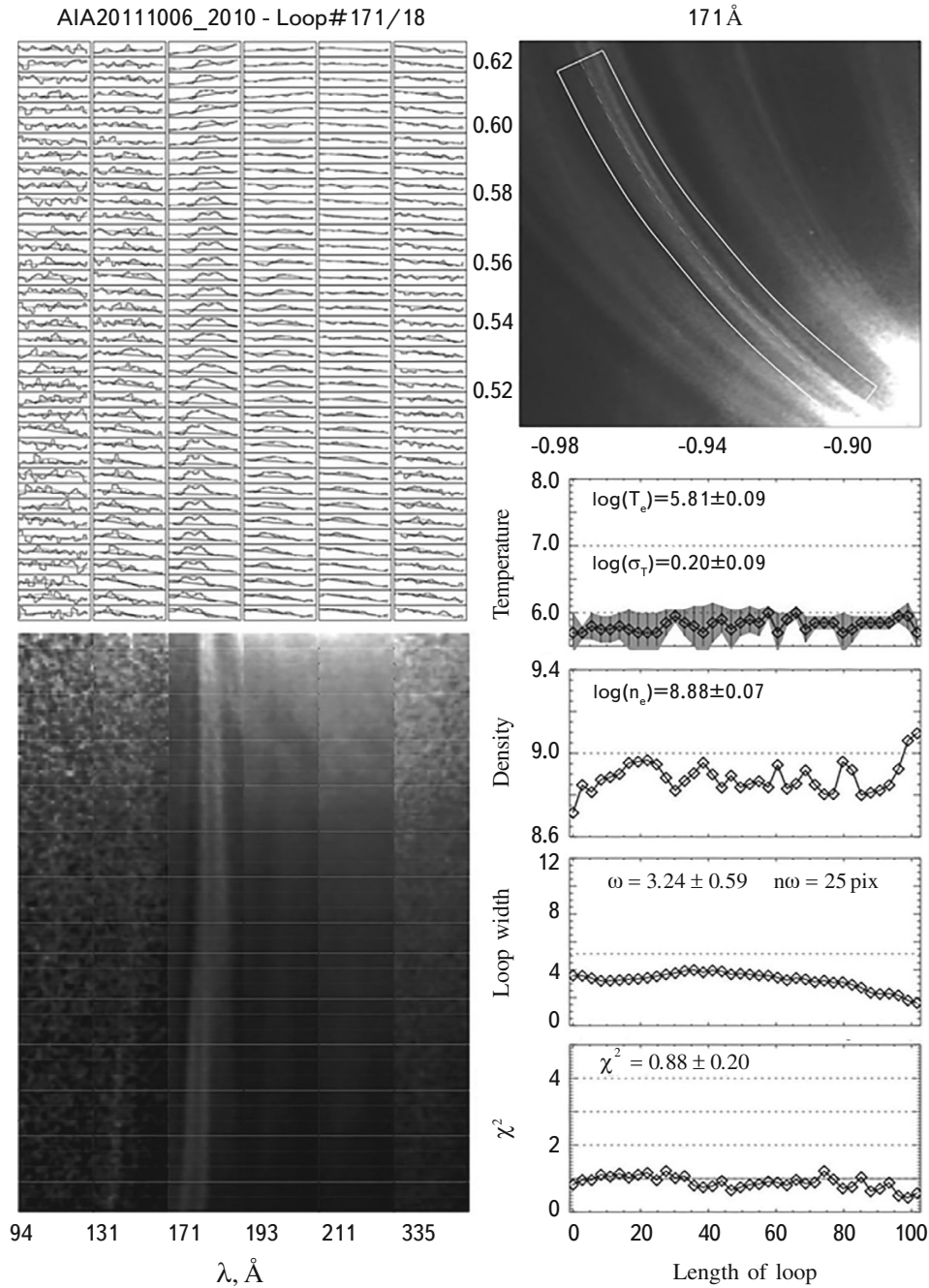

Fig. 3. (Upper right frame) A partial image of an automatically detected profile in the 171 Å filter. (Upper left) Profiles of the transverse cross section flow. (Lower left) Segments of a loop of a cut at different wavelengths. (Four bottom frames on right) The best values of the peak DEM temperature, electron density, and loop width, as well as a $\chi^2$ estimate for this approximation.

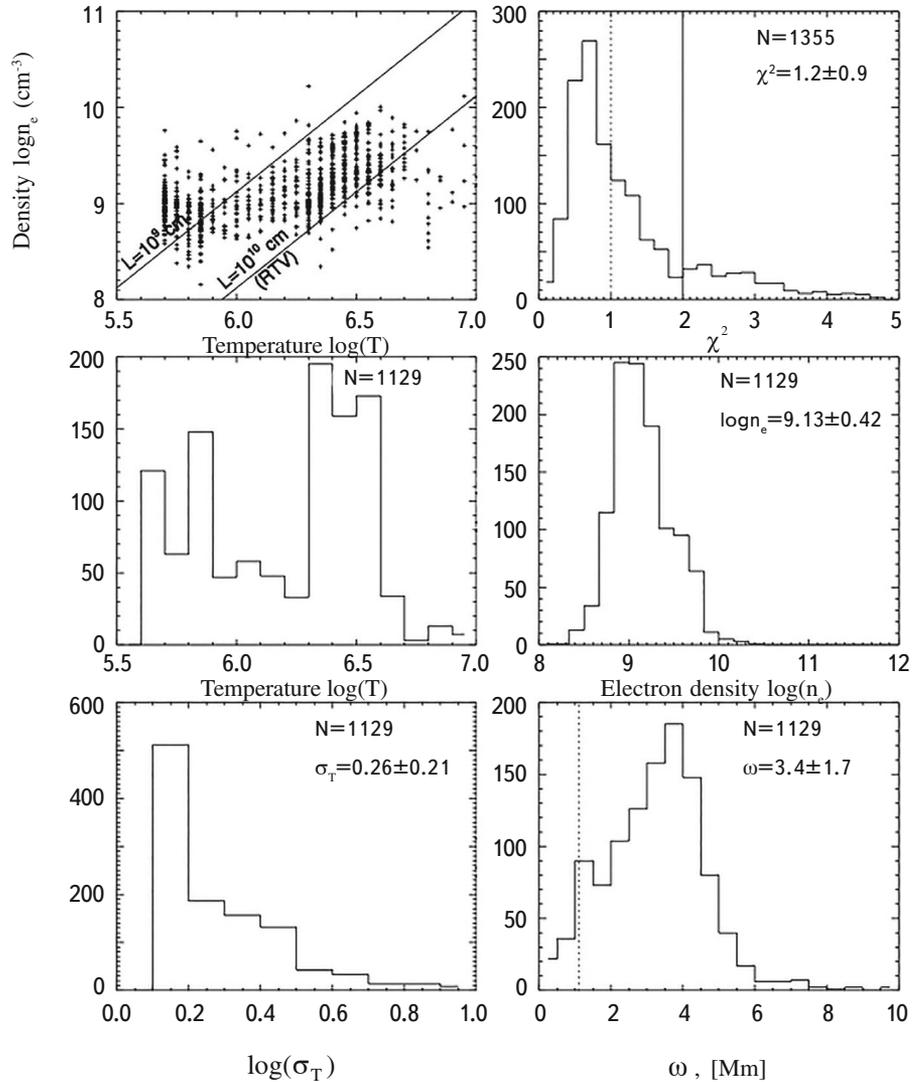

Fig. 4. Statistics of the automated DEM set in 240 tracked coronal loops broken up into 1355 subsegments at the onset of a coronal rain (top right). The upper left corner shows the electron density *vs.* electron temperature with a Rosner, et al., scaling law [25]. Middle left, temperature distributions of the coronal loop. Bottom left: the distribution of the widths of the Gaussian temperature. Middle right: distribution of $\chi^2$ for single-Gaussian approximations. Bottom right: distribution of the loop width.

An automated set of DEM in 240 tracked segments of a coronal loop at the onset and end of a coronal rain is shown in Figs. 4 and 5. The statistics indicate a peak loop temperature of log(T)=6.3-6.6 K (middle frames on the left) and an electron density of about log($n_e$)=9.13±0.42 cm$^{-3}$ in it (middle frame on the right), while the Gaussian width of the DEM temperature distribution is $\sigma_T = 0.26 \pm 0.21$ (lower left frame) and the distribution of the width of the loop reaches $\omega = 3.2 \pm 1.7$ Mm. The measured emission as a function of temperature obeys the Rosner, Tucker, and Vaiana (RTV) scaling law [25], which establishes the relationship between the maximum temperature, pressure, and length of a loop (the diagonal strips in the upper left frames of Figs. 4 and 5 show the predicted RTV scaling law for coronal loops with lengths of 10-100 Mm).

The pressure $p$ of the gas is given by the ideal gas law as

$$p = n_e k_B T, \qquad (1)$$

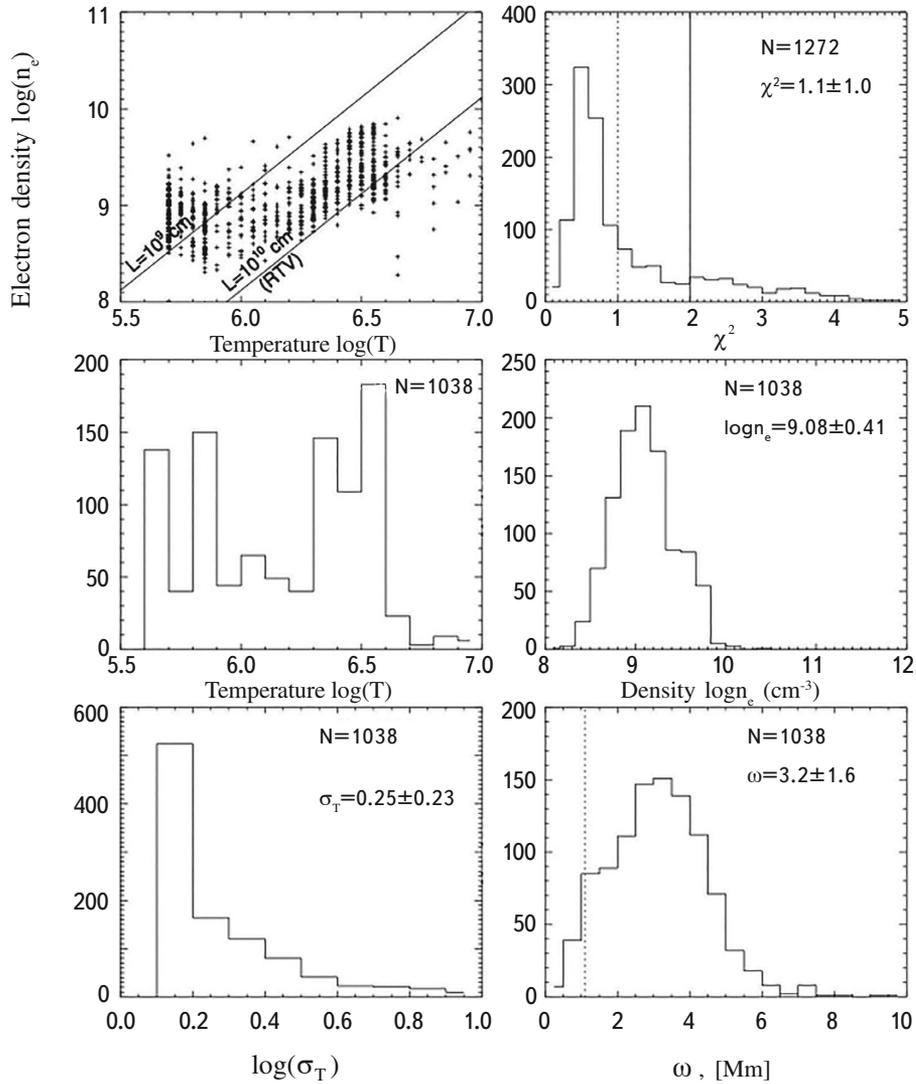

Fig. 5. Same as in Fig. 4, but for the end of the coronal rain.

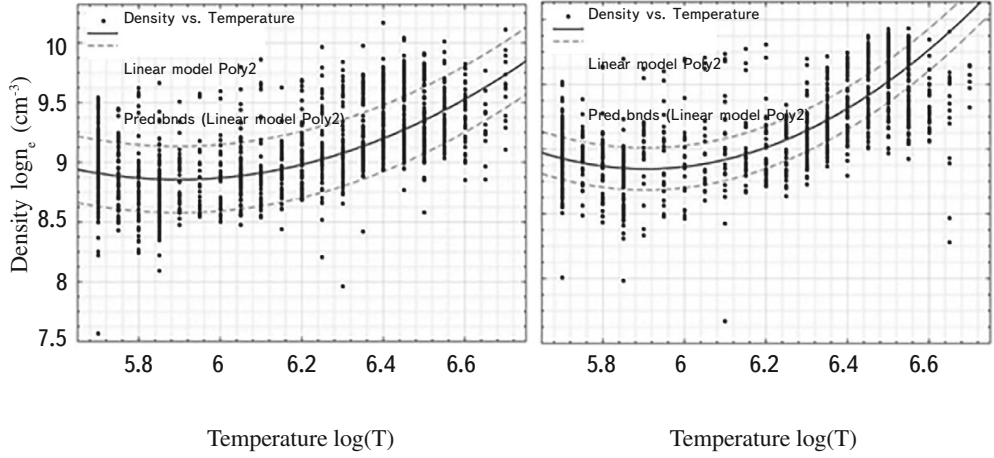

Fig. 6. Electron density *vs.* electron temperature at the onset of a coronal rain with a polynomial approximation (smooth curve, left frame) and a 95% confidence level (dashed curves, left frame). The right frame shows the electron density *vs.* temperature at the end of the coronal rain with a polynomial fit (smooth curve, right frame) and 95% confidence levels (dashed curves, right frame).

where $k_B$ is the Boltzmann constant, $n_e$ is the electron number density, and $T$ is the temperature.

Polytropic processes obey the equation

$$p^{1-\gamma}T^\gamma = \text{const}, \qquad (2)$$

where $\gamma$ is the ratio of the specific heats, which usually lies within the range $\gamma \in [1; 5/3]$. Equations (1) and (2) can be used to obtain a theoretical relationship between the electron density and temperature in polytropic processes:

$$(\gamma - 1)\log(n) = \log(T) - \log\frac{\text{const}}{k_B^{1-\gamma}}. \qquad (3)$$

On the other hand, the statistical significances of the emission and temperature measures in many coronal loops make it possible to plot the relationship between the electron density and temperature. Figure 6 shows the electron density *vs.* electron temperature with a corresponding second order polynomial fit at the onset and end of the coronal rain. At the end of the coronal rain the adiabatic index is estimated to be $\gamma = 1.3\pm0.06$; as expected, this is between isothermal ($\gamma = 1$) and adiabatic ($\gamma = 5/3$) processes. In addition, at the onset of the coronal rain the estimated adiabatic index is $\gamma = 2.1\pm0.11$, which means that the process is not quasiadiabatic as the coronal rain begins to form. The higher value of $\gamma$ probably is evidence of catastrophic cooling.

## 4. Conclusions

We have analyzed the measurement of temperature and emission in the sun's corona near an active region with the automatic AIA/SDO DEM code [22]. 240 coronal loops were automatically detected and analyzed using a DEM code in six EUV channels during a coronal rain that took place from 20:10 UT on October 6 through 02:10 UT on October 7, 2011. The automatically determined mean coronal loop temperature, temperature width, electron density, and loop width were $\log(T) = 6.3\text{-}6.6\,K$, $\sigma_T = 0.26 \pm 0.21$, $\log(n_e) = 9.13 \pm 0.42\,cm^{-3}$, and $\omega = 3.2 \pm 1.7$ Mm, respectively. A contour plot of the electron temperature density distribution compared with the electron distribution of the automatically detected segments of a coronal loop can be used to estimate the adiabatic index at the onset and end of a coronal rain. We found $\gamma = 1.3 \pm 0.06$ at the end of the coronal rain, i.e., at 02:10 UT on October 7, a value between isothermal and adiabatic indices, so it may correspond to a stable coronal plasma. At the onset of the rain (20:10 UT on October 6), the index was estimated to be $\gamma = 2.1 \pm 0.11$, which indicates a slight thermal instability, apparently related to extremely rapid cooling which triggers the mechanism for formation of the coronal rain.

This work was supported by the Shota Rustaveli National Science Foundation (SRNSF) [PhDF2016_147] and grant No. 217146.

## REFERENCES

1. S. K. Antiochos, P. J. MacNeice, D. S. Spicer, et al., Astrophys. J. **512**, 985 (1999).
2. P. Antolin, K. Shibata, and G. Vissers, Astrophys. J. **716**, 154 (2010, P. Antolin, E. Verwichte, Astrophys. J. **736**, 121 (2011).
3. P. Antolin and L. Rouppe van der Voort, Astrophys. J. **745**, 152 (2012).
4. P. Antolin, G. Vissers, and L. Rouppe van der Voort, Sol. Phys. **280**, 457 (2012).
5. G. B. Field, Astrophys. J. **142**, 531 (1965).
6. E. N. Parker, Astrophys. J. **117**, 431 (1953).
7. C. J. Schrijver, Solar Phys. **198**, 325 (2001).
8. K. Murawski, T. V. Zaqarashvili, and V. M. Nakariakov, Astron. Astrophys. **533**, A18, 5 (2011).
9. T. Van Doorsselaere, N. Wardle, G. Del Zanna, et al., ApJL, 727:L32 (4pp) (2011).
10. A. Fludra and J. Sylwester, Solar Phys. **105**, 323 (1986).
11. J. W. Brosius, J. M. Davila, R. J. Thomas, et al., Astrophys. J. Suppl. **106**, 143 (1996).
12. V. Kashyap and J. J. Drake, Astrophys. J. **503**, 450 (1998).
13. M. J. Aschwanden and L. W. Acton, Astrophys. J. **550**, 475 (2001).
14. M. J. Aschwanden and P. Boerner, Astrophys. J. **732**, 81 (2011).
15. M. JAschwanden, P. Boerner, C. J. Schrijver, et al., Solar Phys. 283:5-30 (2013).
16. M. J. Aschwanden and R. W. Nightingale, Elementary loop structures in the solar corona analyzed 387 from TRACE triple-filter images. Astrophys. J. **633**, 499-517 (2005).

17. M. J. Aschwanden and C. J. Schrijver, Astrophys. J. Suppl. **142**, 269 (2002).
18. M. J. Aschwanden, Physics of the Solar Corona (Praxis/Springer, Chichester/New York) (2004).
19. M. J. Aschwanden, Sol. Phys. **262**, 235 (2010a).
20. M. J. Aschwanden, Sol. Phys. **262**, 399 (2010b).
21. M. J. Aschwanden, Sol. Phys. **262**, 399 (2010).
22. M. JAschwanden, Sol. Phys. 283:5-30 DOI 10. 1007/s11207-011-9876-5 (2013).
23. J. R. Lemen, A. M. Title, D. J. Akin, et al., Sol. Phys. **275**, 17 (2012).
24. Z. Vashalomidze, V. Kukhianidze, T. V. Zaqarashvili, et al., Astron. Astrophys. **577**, id. A136 (2015).
25. R. Rosner, W. H. Tucker, and G. S. Vaiana, Astrophys. J. **220**, 643 (1978).